\documentclass{emulateapj}

\usepackage{graphicx}

\begin{document}

\title{Tidal debris from high-velocity collisions as fake dark galaxies:\\ A numerical model of VirgoHI21}

\author{Pierre-Alain Duc \& Frederic Bournaud} 
\affil{Laboratoire AIM, CEA/DSM - CNRS - Universit\'e Paris Diderot\\ Dapnia/SAp, CEA--Saclay, 91191 Gif-sur Yvette Cedex, France}

\shorttitle{Tidal debris as fake dark galaxies: VirgoHI21}
\shortauthors{Duc \& Bournaud}

\begin{abstract}

High speed collisions, although current in clusters of galaxies, have long been neglected, as they are believed to cause little damages to galaxies, except when they are repeated, a process called ``harassment". In fact, they are able to produce faint but extended gaseous tails. Such low-mass, starless, tidal debris may become detached and appear as free floating clouds in the very deep HI surveys that are currently being carried out. We show in this paper that these debris possess the same apparent properties as the so-called ``Dark Galaxies", objects originally detected in HI, with no optical counterpart, and presumably dark matter dominated. We present a numerical model of the prototype of such Dark Galaxies -- VirgoHI21 --, that is able to reproduce its main characteristics: the one-sided tail linking it to the spiral galaxy NGC 4254, the absence of stars, and above all the reversal of the velocity gradient along the tail originally attributed to rotation motions caused by a massive dark matter halo and which we find to be consistent with simple streaming motions plus projection effects. According to our numerical simulations, this tidal debris was expelled 750~Myr ago during a fly-by at 1100~km~s$^{-1}$ of NGC~4254 by a massive companion which should now lie at a projected distance of about 400~kpc. A candidate for the intruder is discussed. The existence of galaxies that have never been able to form stars had already been challenged based on theoretical and observational grounds. Tidal collisions, in particular those occurring at high speed, provide a much more simple explanation for the origin of such putative Dark Galaxies.

\end{abstract}
\keywords{galaxies: interactions --- galaxies: kinematics and dynamics --- galaxies: individual (NGC 4254,VirgoHI21) }

\section{Introduction}

With the availability of unprecedented deep HI blind surveys, a population of apparently free floating HI clouds without any detected stellar counterpart has become apparent \citep{meyer04,davies04,things,alfalfa1,alfalfa2}. It has been suggested that a fraction of them could be ``dark galaxies", a putative family of objects that would consist of a baryonic disk rotating in a dark matter halo, but that would differ from normal galaxies by being free of stars, having all their baryons under the form of gas. They would thus be ``dark'' in the optical and most other wavelengths, but visible through their HI emission, contrary to pure ``dark matter'' halos. 
Such dark galaxies would be extreme cases of Low Surface Brightness Galaxies (LSBs), a class of objects that have a particularly faint stellar content compared to their gaseous and dynamical masses \citep[e.g.,][]{carignanfreeman88}. The formation of low mass dark galaxies is actually predicted by $\Lambda$-CDM models \citep[e.g.,][]{vandenbosch03,tully05}. \citet{taylor05} provided theoretical arguments against the existence of galaxies that would have remained indefinitely stable against star formation, unless they are of very low mass, at least a factor ten below than that of classical dwarf galaxies. 

If they exist, the dark galaxies are predicted to have a low dynamical mass and HI content. In the Local Group, some possibly rotating, high-velocity clouds were speculated to be dark galaxies \citep{simon04,simon06}. Further away, \citet{davies06} argued that most previous HI blind surveys were not sensitive enough to rule out the existence of dark galaxies. And indeed, while the HIPASS survey failed at detecting HI~clouds without optical counterparts \citep{doyle05}, deeper recent HI observations, in particular with the Arecibo telescope, have revealed a number of dark galaxy candidates \citep{alfalfa2}. Among these free-floating low-mass HI clouds, one object located in the outerskirts of the Virgo Cluster has attracted much attention and discussion: VirgoHI21 \citep[][see Fig.~1]{davies04,minchin05}. Despite a HI mass of only $\sim 10^8$~M$_\sun$, this elongated gaseous structure, mapped with the Westerbork Synthesis Radio Telescope (WSRT) by \citet[][hereafter M07]{minchin07}, exhibits a velocity gradient as large as 220~km~s$^{-1}$ (see Fig.~\ref{fig:pv-vhi21}). {\it Assuming} that the observed HI velocities trace rotation, the inferred dynamical mass would be as large as $\sim 10^{11}$~M$_\sun$. The object shows no optical counterpart, even on deep HST images (M07). With such extreme properties, VirgoHI21 has become the prototype for dark galaxies, although its high dynamical mass is atypical even in models predicting the existence of Dark Galaxies. If real, an object like VirgoHI21 could tidally disturb the galaxies in their neighborhood, as investigated by \cite{karachentsev06}. Actually VirgoHI21 itself lies at about 150~kpc from the massive spiral galaxy NGC~4254 (M~99), to which it is connected by a faint HI~filament. This structure could in principle be a bridge linking the two galaxies and would then appear as a sign of a tidal interaction between them (M07).

\begin{figure*}
\centering
\includegraphics[width=\textwidth,angle=0]{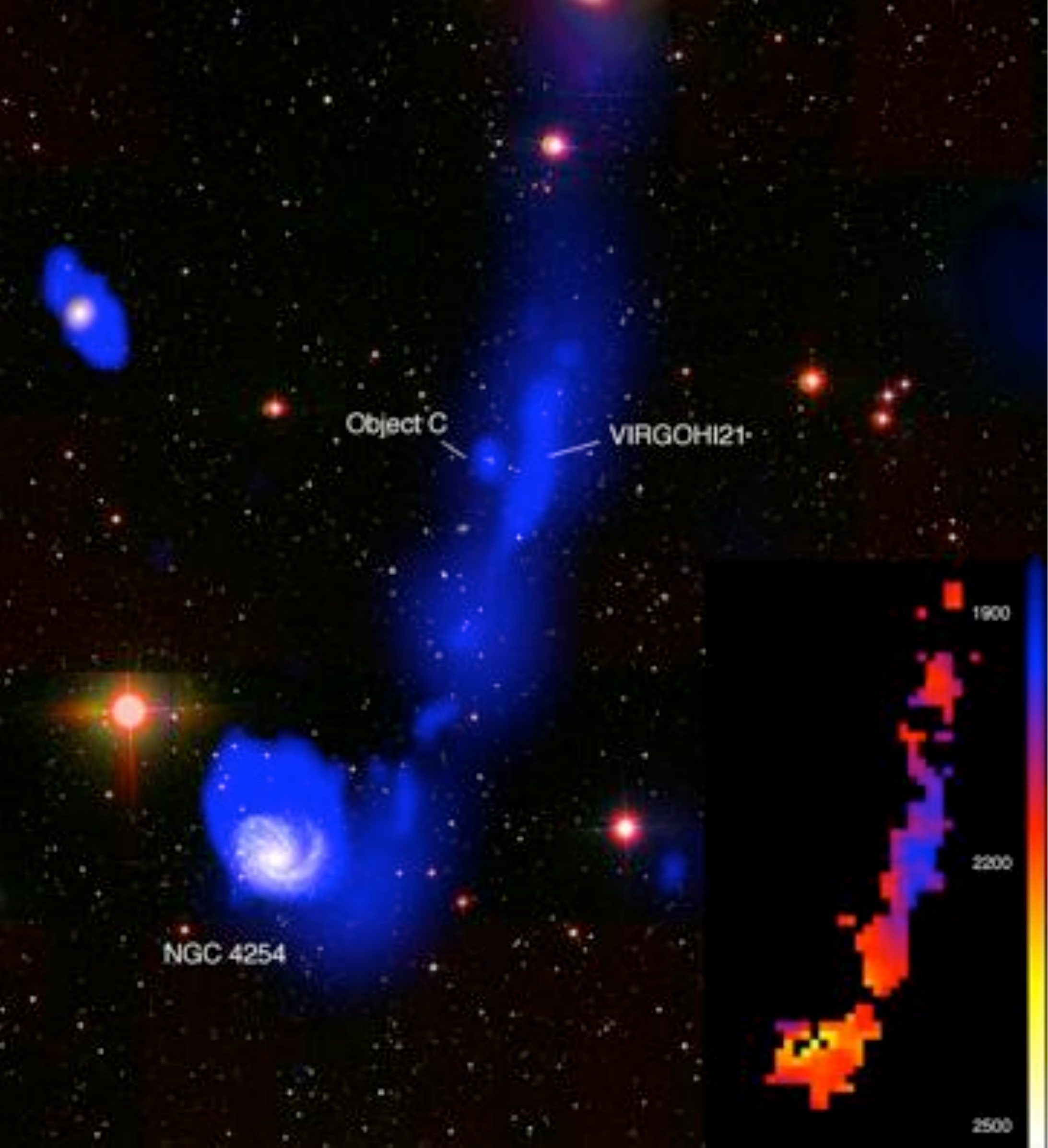}
\caption{The system VirgoHI21/NGC 4254. {\it Left} The distribution of the atomic hydrogen is superimposed in blue on a true color optical SDSS image of the field acquired through the WIKISKY.org project. The HI maps actually combine two data set: the observations obtained with the Arecibo telescope as part of the ALFALFA project \citep[courtesy of B. Kent,][]{haynes07}, which are sensitive enough to show the whole extent of the gaseous tail; the observations obtained at WSRT \citep[courtesy of R. Minchin,][]{minchin07} which are not as deep, but have a much better spatial resolution. The HI maps were smoothed and the WRST data have been deconvolved by the elongated telescope beam. {\it Right} Color coded first moment, velocity, field of the HI tail as mapped by Arecibo. The observed velocity range in km~s$^{-1}$ is indicated to the right.} \label{fig:vhi21}
\end{figure*}

\begin{figure}
\centering
\includegraphics[width=\columnwidth,angle=0]{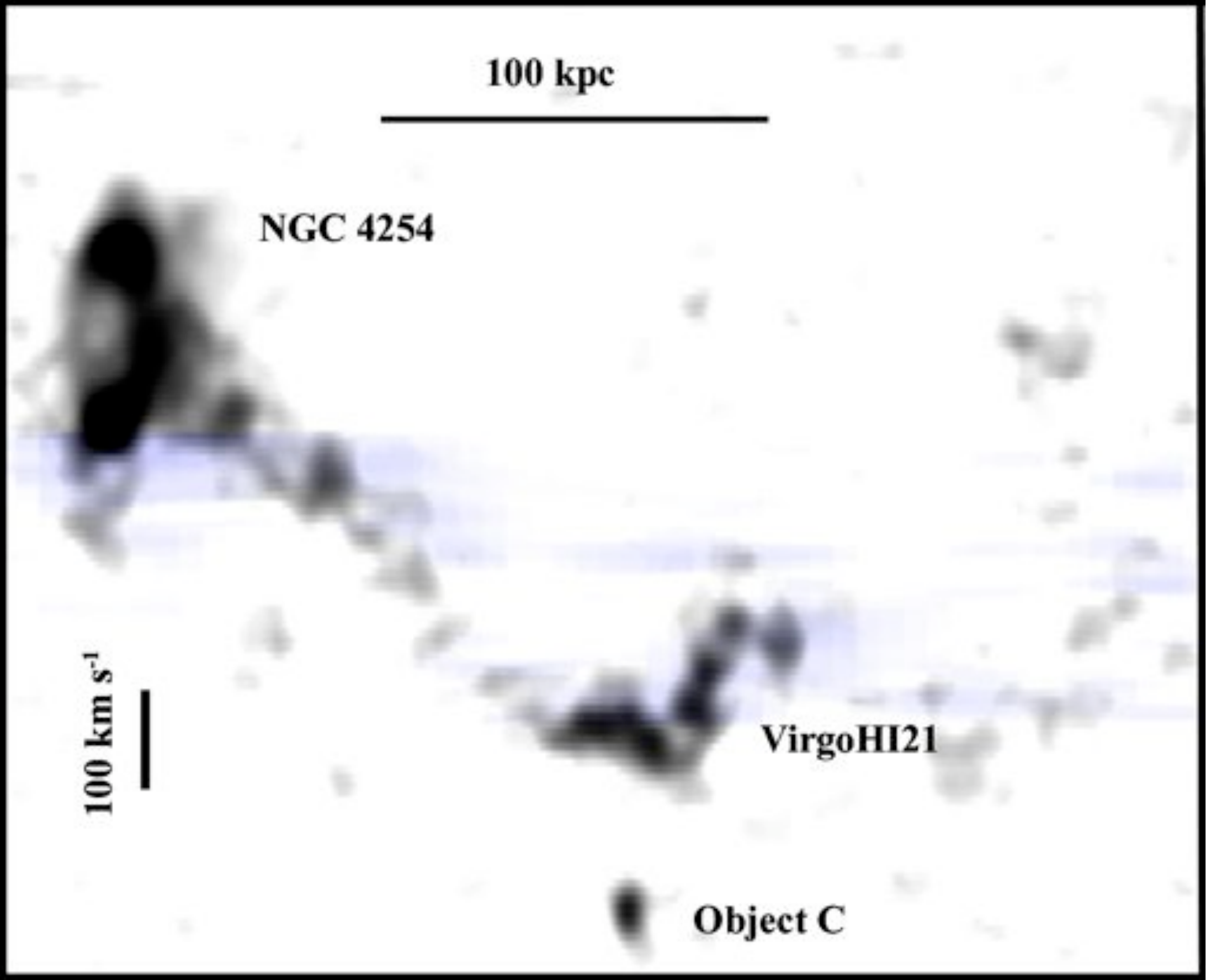}
\caption{Observed Position (x-axis) Velocity (y-axis) diagram along the position angle 22 degrees corresponding to the main direction of the HI bridge. As for Figure~\ref {fig:vhi21}, the WSRT (gray) and Arecibo (light blue) data set have been superimposed to show all available information.}\label{fig:pv-vhi21}
\end{figure}

However, starless isolated gas clouds, showing a large velocity spread, are not necessarily genuine dark galaxies. Ram pressure can strip gas away from spirals in the vicinity of clusters, a process that does not affect stars. Interaction with an external field, for instance that of another galaxy, can expulse large amounts of material from the disk in the form of gas-rich tidal tails and debris. In that vein, \citet[][hereafter B05]{bekki05}  have suggested that interactions between flying-by (i.e. interacting without merging) galaxies orbiting in a potential well similar as the one produced by the Virgo Cluster  form tidal tails that after some time can resemble isolated gas clouds containing little stars. Furthermore tidal tails are the place of large streaming motions, as shown for instance by the observations of the merger prototype NGC~7252 by \citet{hibbard94} and the models of \citet{bournaud04} and B05. The remaining HI structures of galaxy collisions can thus not only appear as isolated HI clouds, but also exhibit large velocity gradients that mimic those expected for rotating disks within dark matter haloes. 

In these conditions, serious doubts have raised whether VirgoHI21 is really a dark galaxy and not simply the result of a tidal interaction or of an harassment process in the Virgo Cluster (B05). They have recently been reinforced by the publication by \citet{haynes07} of a deep HI map of the field, obtained with the Arecibo Telescope as part of the ALFALFA survey \citep{alfalfa1,alfalfa2}. It reveals that VirgoHI21 actually lies within an even larger HI structure that extends further to the North, in the opposite direction of NGC~4254. This further suggests that the thin and long HI feature is in fact a tidal tail emanating from the spiral and that VirgoHI21 is just a denser cloud within it and thus not a dark galaxy.  On the basis of HI spectra obtained with the  Effelsberg telescope and a numerical model including the cluster ram pressure, \citet{vollmer05} suggested that NGC~4254 has indeed interacted recently with a companion. However this study mostly  focussed on the internal properties of the spiral and in particular the formation of  VirgoHI21 was not modeled. Nevertheless, several arguments against a tidal origin for the HI cloud have been raised  and addressed in detail in M07. The main ones regard the absence of a suitable interacting companion,  the nonexistence of a counter tail -- a feature that is generally present in tidal interactions --,  the total lack of stars in the HI tail/bridge,  and, above all, the remarkable 200~km~s$^{-1}$ velocity gradient associated to VirgoHI21, which seems to be reversed and amplified with respect to the large scale velocity field along the rest of the HI structure. 

We will show in this paper that most of these criticisms actually apply to low-velocity encounters and not to the high-velocity ones, which are common in the cluster environment. High-velocity collisions have been neglected so far because they seem to cause little disturbances to stellar disks unless they are very numerous and participate to an harassment process \citep[e.g.,][]{moore96}. To further investigate the role of high speed collisions in the formation of tidal debris, especially the gaseous ones, we have carried out a series of numerical simulations. We illustrate their impact showing a numerical model reproducing the morphology and kinematics of VirgoHI21.

The numerical simulations are presented in Section~2. In Section~3, we compare the properties of tidal tails formed in high- and low-velocity galaxy encounters. In Section~4, we present the model which best fits VirgoHI21. Its actual nature is discussed in Section~5. Our conclusions and the implications for the detection of dark galaxies are summarized in Section~6.

\section{Numerical simulations and parameters}

\subsection{Code}
We model the encounter of spiral galactic disks and dark matter haloes using a particle-mesh sticky-particle code \citep[e.g., ][]{BC02, BC03}. The gravitational potential is computed on a Cartesian grid with a softening length of 150~pc. Stars, gas, and dark matter haloes are modeled with one million particles each. The interstellar gas is modeled using sticky-particle parameters $\beta_r=\beta_t=0.7$ \citep[defined as in][]{BC02}. Star formation is described by a local Schmidt law. At each timestep, the probability for each gas particle to be transformed into a stellar particle is proportional to the local gas density to the exponent 1.4 \citep{kennicutt98}, the proportionality factor being computed to provide a star formation rate of $2$~M$_{\sun}$~yr$^{-1}$ in the initial spiral disk of the target galaxy,  which ensures a realistic gas-consumption timescale of 4--5~Gyr. 

A star formation threshold at gas surface density above 3~M$_\sun$~pc$^{-2}$ is applied. The choice of a  threshold  lower than the average value of typical spirals \citep{martinkenn01}  but within the range of values  found by  \citet{guiderdoni87} for anemic Virgo spirals,  is a conservative one: it ensures that the absence of stars in the VirgoHI21-like tidal debris  is a robust result and not an artefact caused by an artificially  high threshold. In fact, the average gas surface density of VirgoHI21 is estimated to 2~M$_\sun$~pc$^{-2}$, given its HI mass of $3\times10^7$~M$_\sun$ distributed over about $14 \times 1$~kpc (M07). It is higher in its densest regions, but not reaching the critical value for the onset of star-formation, just like in our model.

\subsection{Initial conditions}
\subsubsection{Target galaxy}
The target galaxy in all our simulations is similar to NGC~4254 (the one connected to VirgoHI21 by a tidal bridge), which is a typical spiral galaxy with a circular velocity of 190~km~s$^{-1}$. Unless specified otherwise, the physical parameters for this galaxy were taken in the {\tt HyperLeda} extragalactic database\footnote{\tt http://leda.univ-lyon1.fr}. 

The stellar disk is truncated at a radius 13~kpc (compatible with the $R_{25}$ optical radius) with initially a \citet{toomre63} profile of scale-length $a=4$~kpc. The gas disk has an initial radius of 30~kpc, with a flat distribution from $r=0$ to r$=25$~kpc and a linear decrease from 25 to 30~kpc. 
This extent is compatible with the presently observed HI disk, taking into account the fact that  the outer parts are stripped during the interaction. The  final HI radius in our model is about 25~kpc, while \citet{phookun93} detect HI up to radii of 20~kpc and the deeper observations by \cite{haynes07} show HI detection up to 25~kpc.  Note that an HI disk truncated at 25 kpc already forms the main HI bridge and the VirgoHI21-like cloud, whereas a slightly larger one is required to account for the 
most recent HI data showing an extension North of  VirgoHI21.

Both the stellar and gaseous disks are initially set up as \citet{toomre63} disks with a radial scale length of 4~kpc for stars and 8~kpc for gas. A central bulge (Plummer profile of scale-length 1~kpc) is implemented, with a bulge:disk mass ratio of 0.25. The dark matter halo is modeled with a softened isothermal sphere with a mass density profile given by:
\begin{equation}
\rho(r)=\frac{\sigma^2}{2 \pi G (r^2+r_c^2)}
\end{equation}
The sphere is truncated at $r=100$~kpc and we use a core radius $r_c=4$~kpc. This profile provides an extended flat rotation curve and was found by \citet{DBM04} to fairly reproduce the distribution of gas along tidal tails when compared to real systems. 
The dark halo mass within the optical radius (13~kpc) is $8.5 \times 10^9$~M$_{\sun}$ (dark-to-visible mass ratio 0.6). Within the outer edge of the initial HI disk at 30~kpc, the halo mass is $2.2 \times 10^{11}$~M$_{\sun}$ (dark-to visible ratio 1.6). The total mass of the dark halo up to its truncation radius at 100~kpc is $9 \times 10^{11}$~M$_{\sun}$. The total dark-to-visible ratio, 7.5:1, is compatible with those favored by \citet{dubinski1996} to form long tidal tails.
 The initial gas mass fraction in the disk is 7\%, compatible with a present-day gas fraction of about 5 percent after star formation. The mass of each component is computed to get a circular velocity $V_\mathrm{circ} \simeq $190~km~s$^{-1}$ at $r=15$~kpc, as observed in NGC~4254; this give a stellar mass of $1.2 \times 10^{11}$~M$_\sun$ and initial gas mass of $8.5 \times 10^9$~M$_\sun$.

\subsubsection{Interloper galaxy and orbital parameters}

The modeled interloper galaxy is gas-free. This choice is made for simplicity and CPU-time efficiency, because the gas disk of the interloper can disturb the target disk only for close encounters and/or low-velocity encounters with coplanar disks, when the two gas disks overlap and shock. This is not the case for the high-velocity fly-bys studied here. In particular the pericenter distance in our model for VirgoHI21 (see Sect.~\ref{sect:virgohi}) is 59~kpc with an orbital plane different from the target disk plane, making any overlap of the gas disks unlikely. We chose to model the interloper as a spiral galaxy\footnote{an elliptical galaxy would produce a quite similar tidal field on the target galaxy, especially during distant encounters without mergers.} with the same mass proportions and relative sizes for the bulge, disk, and dark matter halo. All sizes are scaled as the square root of the mass relatively to the target galaxy, which keeps the central density constant.

We performed a number of simulations aimed at comparing the formation of tidal tails in high and low velocity encounters. In all the seven runs we have carried out, the orbit is direct, i.e. the orbital momentum of the interloper w.r.t. the target has the same direction as the spin of the target disk. The orbital parameters, which are given in Table~\ref{tab:param} for all runs, are: \\
-- the initial velocity $V_\infty$ at an infinite distance; the velocity at the beginning of the simulations is computed with the assumption that the dynamical fraction is negligible before the beginning of the simulation.\\
-- the impact parameter $b$, which is the distance at which the two galaxies would cross if they had linear trajectories along their initial velocity.\\
-- the actual pericenter distance $R_\mathrm{P}$ and velocity at the pericenter $V_\mathrm{P}$. These quantites are measured from the simulation.\\
-- the inclination of the orbit plane w.r.t. the target galaxy disk plane $i$. A coplanar encounter corresponds to $i=0$.\\
-- the ratio of the interloper-to-target total masses $M_i$.\\

\begin{table}
\begin{center}
\caption{Run parameters}\label{tab:param}
\begin{tabular}{cccccccc}
\tableline\tableline
Run && $V_\infty$ & $b$ & $i$ & $R_\mathrm{P}$ & $V_\mathrm{P}$ & $M_i$ \\
\tableline
1 && 900 & 90 & 25 & 67 & 1050 & 1 \\
2 && 230 & 125& 25 & 72 & 320 & 1 \\
3 && 900 & 65 & 35 & 41 & 1145 & 1 \\
4 && 230 & 85 & 35 & 49 & 355 & 1 \\
5 && 900 & 50 & 40 & 33 & 1195 & 1 \\
6 && 230 & 70 & 40 & 27 & 380 & 1 \\
\hline
7 && 900 & 63 & 41 & 59 & 1125 & 1.5 \\

\tableline
\end{tabular}
\end{center}
\end{table}

\section{Tidal tails in high velocity encounters}\label{sect:highvel}

We compare here the properties of tidal tails formed in low and high velocity encounters. All these simulations are for direct orbits, with high velocity ($V_\infty =900$~km~s$^{-1}$) for runs 1-3-5, and low velocity ($V_\infty =230$~km~s$^{-1}$) for runs 2-4-6, the other parameters being unchanged. The impact parameters $b$ were chosen to provide comparable pericenter distances $R_\mathrm{P}$ for each simulation pairs 1-2, 3-4, 5-6 (although the exact pericenter distance depends on dynamical friction and cannot be exactly predicted). We compare high- and low-velocity encounters at (roughly) fixed pericenter distance and inclination for each pair.

\begin{figure*}
\centering
\includegraphics[width=4.4in]{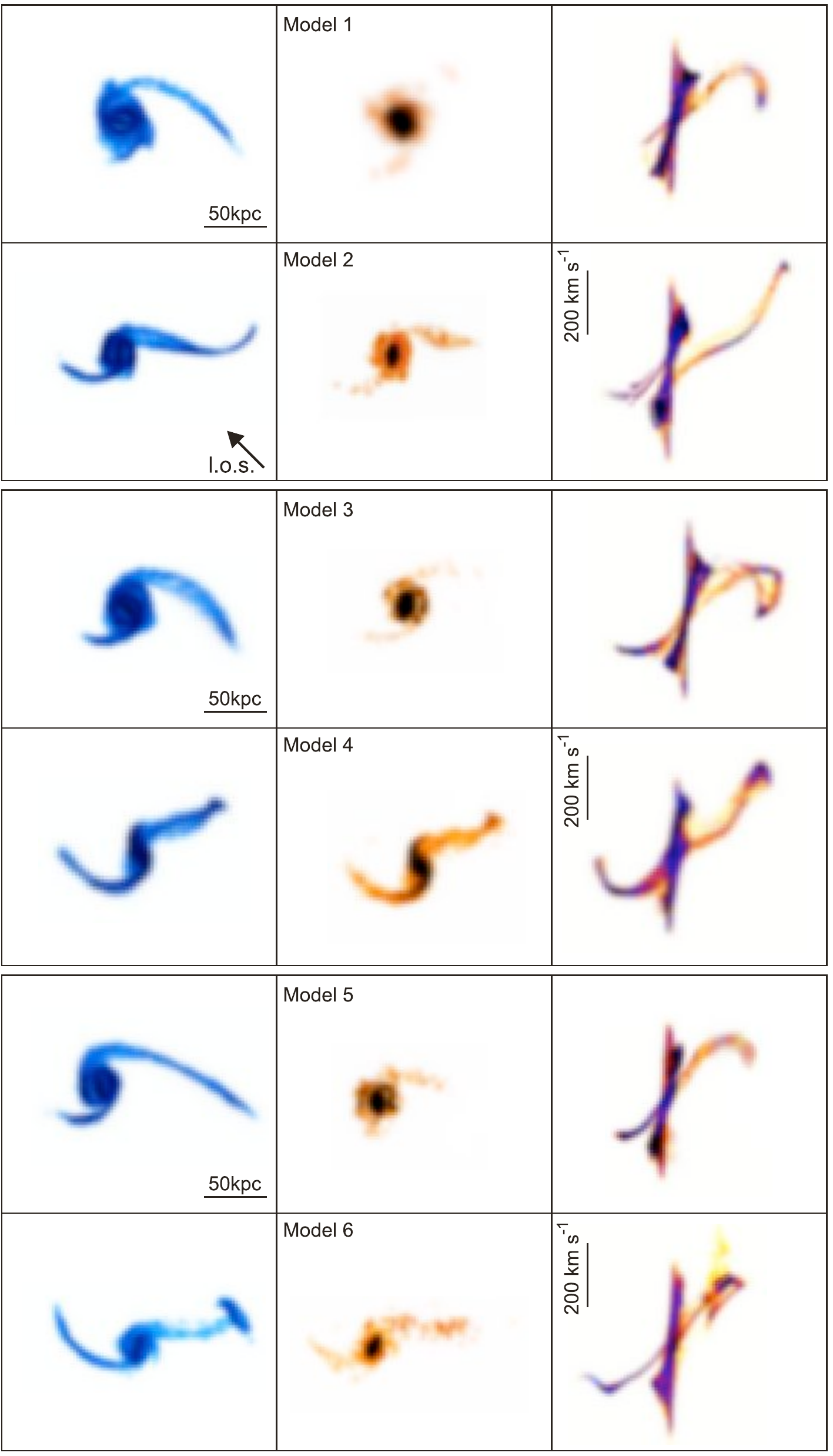}
\caption{A comparison between the effects of low and high velocity encounters on the gaseous component (pictures to the left), on the stellar component (pictures in the middle) and on the velocity profile (pictures to the right). The parameters of the six numerical simulations presented here are listed in Table~\ref{tab:param}. The high velocity cases correspond to run 1, 3 and 5. 
The arrow shown in the left picture of Model 2 indicates the line of sight used for all models to produce the Position-Velocity diagrams displayed in the right panels.}\label{fig:highvel}
\end{figure*}

In Figure~\ref{fig:highvel}, we present for stars and gas separately the face-on morphology 300~Myr after the pericenter passage. We also show on this figure the Position-Velocity plots of the tails seen edge-on.

From this comparison we infer that:

\begin{itemize}
\item high-velocity encounters with $V_\mathrm{P} \sim 1000$~km~s$^{-1}$ or more can form gaseous tidal tails. The mass of the tails, defined as the mass at more than 1.5 times the initial HI disk radius, is on average 17\% of the gas mass for the low-velocity cases, and 9\% for the high-velocity cases (both measured 300~Myr after the pericenter). 
\item the gaseous tails formed during high-velocity encounters can be as long as those formed in low-velocity encounters.
\item the counter-tail (opposite to the main tail) is fainter and shorter, hence falling back more rapidly onto the parent spiral disk. One-tailed systems can hence be obtained more easily for high-velocity encounters, in a few $10^8$~yr.
\item the stellar content is lower for high velocity encounters: the average stellar mass fraction in the tails for low-velocity cases is 18\%, while it is 6\% for high-velocity cases, for the same progenitor spiral galaxy. Most stars, if not all, remain in the parent disk, which is more moderately disturbed. 
\end{itemize}

High-velocity encounters are thus in a sense less efficient than lower-velocity encounters to form tidal tails: the stellar tails and gaseous counter-tail are faint or absent. Still, they do manage to form long tidal tails of pure gas.
Indeed, even if the linear velocity of the interloper is well above the circular velocity of the target disk, its angular velocity $\Omega_\mathrm{interloper} = V_\mathrm{P}/R_\mathrm{P}$ is typically 1000 ~km~s$^{-1} \times sin(i) / 50$~ kpc (projected in the target disk plane), i.e. roughly similar to the disk material angular velocity, $V_\mathrm{circ}/R_\mathrm{disk}$ equal to 200~km~s$^{-1}$ / 15~kpc. These resonant velocities enable the formation of tidal tails even for high-velocity encounters for which the close interaction does not last long.

Note on Figure~\ref{fig:highvel} that higher-velocity encounters affect only the very outer disk, leaving most of the initial disk intact after the interaction. Lower-velocity interactions affect the target disk more deeply inwards, which leaves a less extended gas disk remnant, as visible on the snapshots. This explains why the tails are more massive in the latter case, and contain more stars from the innermost regions of the parent galaxy.

We also note on Figure~\ref{fig:highvel} that the total velocity excursion of a tidal tail $\Delta V$ is not correlated with the encounter velocity $V_\mathrm{P}$, but is rather of the order of the parent disk circular velocity. Hence, contrary to what intuition may suggest,  velocity spreads along tidal tails of only 200~km~s$^{-1}$ can be caused by collisions at speeds higher than 1000~km~s$^{-1}$. The velocity spread of tidal tails is thus essentially influenced by the total mass of the progenitor. Furthermore, all tidal tails may exhibit apparent changes of sign in their projected velocity gradient. This has been shown for low-velocity encounters by \citet{bournaud04} but is still true for high velocity encounters. These velocity gradients result from streaming motions and differential rotations along the tidal tails, but do not trace any rotation and cannot be interpreted in terms of dynamical mass.

 In summary, high-velocity encounters may create low-mass but long tidal  tails. This result goes against common belief but, as a matter of fact, very few works had previously investigated in detail  this type of fly-bys.  In their pioneering modeling work,   \cite{eneev73} had shown that  encounters at the moderate velocities of 400~km~s$^{-1}$ were able to create tidal arms. They however noted that the ejection of material should decrease when the  duration of interaction gets smaller, and thus when the encounter velocity increases.  B05 did consider  fly-bys, but in their model the formation of tidal tails  was helped by the cluster tidal field. 
We have shown here that high-velocity encounters {\it  alone} can produce long gaseous tidal tails provided that  the HI disk in which they form is initially large enough. It is known to be  2--3 times more extended than stellar disks \citep[e.g.,][]{robertshaynes94} but this extent is often neglected in numerical models that aim at reproducing the internal structures of interacting systems.  Such high-velocity encounters indeed do not affect much the stars and ISM in the inner regions.

\section{A model for VirgoHI21}
\label{sect:virgohi}

We now more specifically investigate whether the HI cloud VirgoHI21 could be part of an HI tidal tail emanating from the spiral galaxy NGC~4254 that would have formed during a high-velocity encounter. As pointed out by M07, the absence of a nearby massive companion makes the hypothesis of a slow encounter very unlikely. 

\begin{figure*}
\centering
\includegraphics[width=\textwidth]{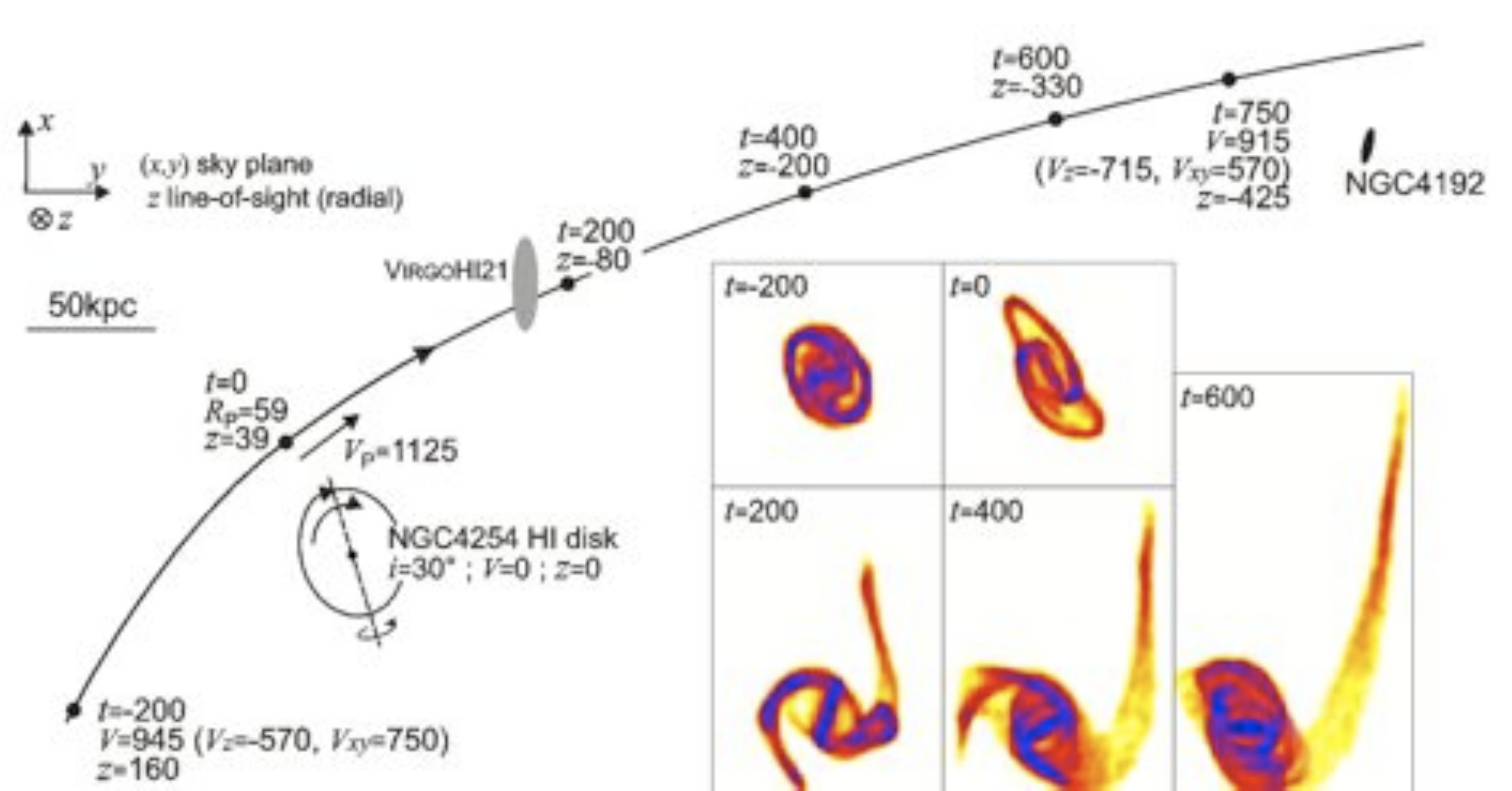}
\caption{Model 7, aimed at reproducing the real system VirgoHI21/NGC 4254. The trajectory of the intruder in the plane of the sky is indicated. It is labeled with the principle parameters of the encounter: time $t$, velocity $V$, position along the line of sight, $z$. The target galaxy is fixed at $V=0$ and $z=0$ in this frame. The insets show snapshots of the simulated gas as a function of time, also projected in the sky plane. Snapshots at t$=750$~Myr are shown in Figure~\ref{fig:vhimodel}.}\label{fig:orbit}
\end{figure*}

\begin{figure}
\centering
\includegraphics[width=\columnwidth]{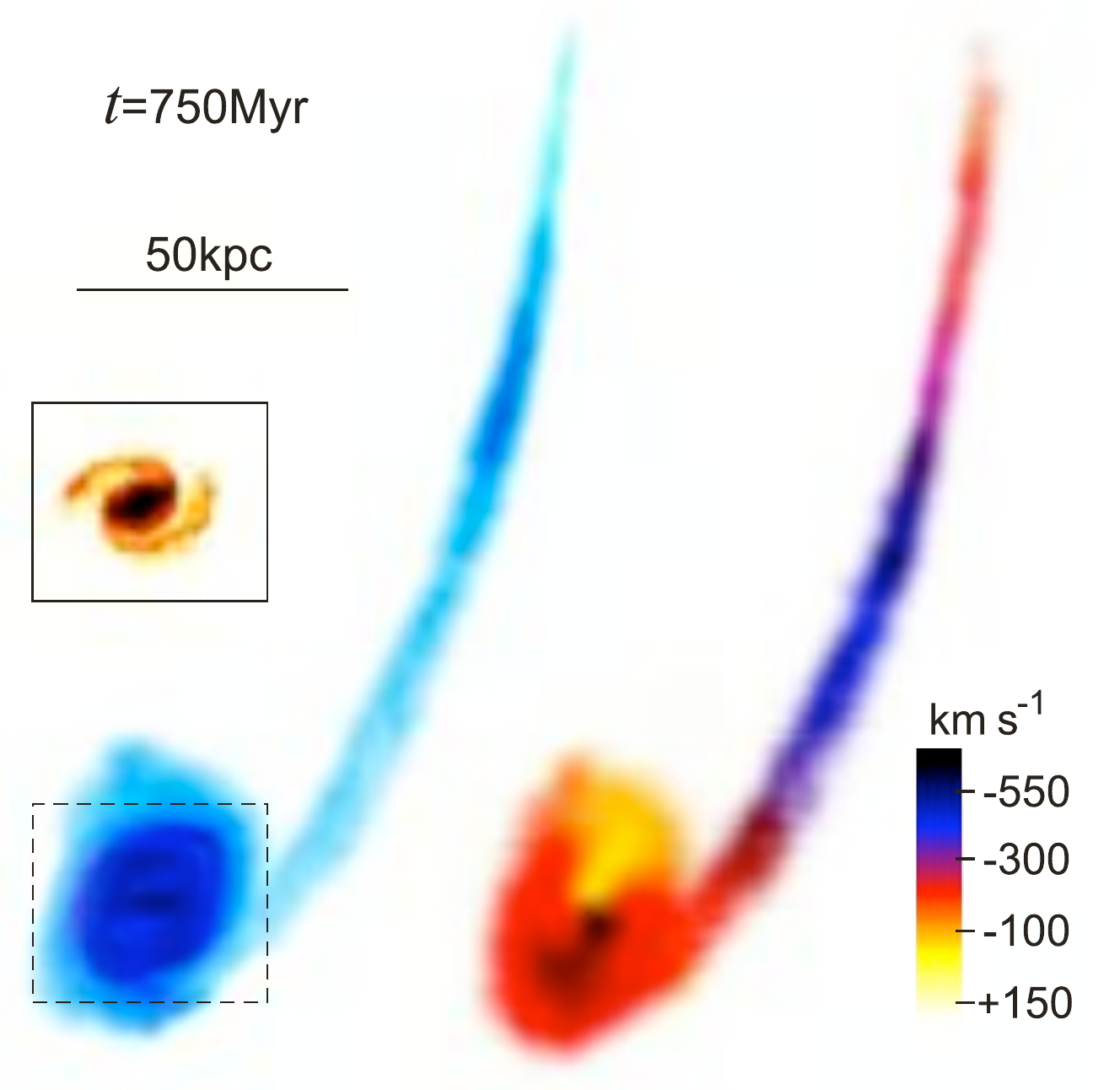}
\caption{Model 7 at $t=$750 Myr when it best reproduces the actual morphology and velocity field of the system VirgoHI21/NGC 4254. The simulated gas distribution is shown to the left (in blue). The inset above shows the simulated stellar component. The color-coded simulated velocity field of the gas is presented to the right. }\label{fig:vhimodel}
\end{figure}

Our numerical model showing the best match with VirgoHI21 -- Model 7 in Table~\ref{tab:param} -- is presented on Figure~\ref{fig:orbit}. The flying-by object has a mass 1.5 times that of NGC~4254, had an initial velocity of 900 km~s$^{-1}$ and a pericentric distance of 59~kpc. The tangential orbital plane at the pericenter makes an angle of 41 degrees with the target disk plane. For simplicity, {\it all} maps in this section (morphology and kinematics) are projected on the plane of the sky as for the observations; note that the parent spiral disk is inclined by 35 degrees with respect to it. Snapshots are shown on Figure~\ref{fig:orbit} up to $t=600$~Myr after the pericenter passage. Figure~\ref{fig:vhimodel} presents the gas and star morphology and the gas velocity field at $t=750$~Myr, at a time when the model best resembles the observed system, as described below.

\subsection{Morphology}

The morphology of the system at $t=750$~Myr is that of a spiral galaxy with a {\it single} and {\it gas-pure} tidal tail. Indeed the counter-tail seen on Figure~\ref{fig:orbit} was short, and fell back rapidly onto the spiral galaxy. Because the companion is massive, the tidal tail is long, with a total extend to the faintest outermost regions of $\sim$200~kpc. Since the orbit is not coplanar to the spiral disk, the tidal tail is seen with a higher inclination than the parent disk itself. This gives the tail a particularly linear shape.

The gas mass of the tail is $2 \times 10^8$~M$_\sun$ (measured as the mass at $r>35$~kpc) while the total gas mass of the system at the same instant is $6.5 \times 10^9$~M$_\sun$. The simulated tail exhibits a  density peak about 150~kpc away from the parent galaxy center. As explained in Section~\ref{fakedark}, the formation in tidal tails of  sub-structures  that may even be gravitationally bound  is not uncommun.
The denser region gathers a gas mass of $5 \times 10^7$~M$_\sun$ over 20~kpc.
 These values are comparable to that observed for the object known as VirgoHI21: M07 give an HI mass of $3 \times 10^7$~M$_\sun$ over 14~kpc. The fact that the tidal structure extends in our model further out beyond the HI condensation is also consistent with the detection of HI North of VirgoHI21 recently reported by \cite{haynes07}.

The outer tail and in particular the HI condensation are basically free of both old and young stars: the fraction of stars in the baryonic mass present at radii $r>35$~kpc is smaller than 3\%. Old stars from the pre-existing galaxy are only found at the base of the tail, forming a faint extension. As noted in Section~\ref{sect:highvel}, the inner regions of the parent galaxy where the bulk of the stellar population lies are weakly disturbed by high-speed collisions. They may only cause the spiral disk to appear modestly lopsided whereas the lopsidedness is much stronger in the HI component, as actually observed for NGC~4254 \citep{phookun93,vollmer05}. The simulated tail is also devoid of your stars. Indeed, the typical gas density along the tail is a few 0.1~M$_\sun$~pc$^{-2}$, not enabling in situ star formation, even with the rather low star-formation threshold chosen in our model (see Sect.~2).

\subsection{Gas kinematics}

The parent spiral galaxy has a circular velocity and inclination chosen to reproduce NGC~4254, i.e. a rotating disk with its receding side to the NE, with typical projected velocities of $\sim$ 100--150~km~s$^{-1}$. The gas kinematics in the inner regions ($r<15$~kpc) is disturbed by a strong bar. Such a strong bar is not present in NGC~4254, but the bar region (inside its corotation) and the regions affected by the tidal interaction (which is outside the stellar disk and the bar outer resonance) are uncoupled, so that this difference does not affect the modeling of the tidal debris. Note also that ram pressure which may further affect the gas kinematics (see  Section~\ref{rampressure}) has not been taken into account in our model.

As noticed before, the orbital spin inclined w.r.t the line-of-sight gives the tidal structures large excursions on the radial direction. The streaming motions create large velocity gradients along the tail. Because the interloper is not in the target disk plane at its pericenter, the tidal tail has significant streaming motions along the line-of-sight, even if the rotation of its parent disk is close to the sky plane. The simulated velocity field shown on Figure~\ref{fig:vhimodel} has the same fundamental properties as that of VirgoHI21 presented on Figure~\ref{fig:vhi21}:
\begin{itemize}
\item the velocity amplitudes along the whole tidal structure and within the HI cloud are similar to that observed. 
\item the radial velocities in the tail are lower than in the spiral galaxy; this corresponds to tidal material falling back into the spiral, hence approaching towards us.
\item the material at the middle of the tidal tail falls back more rapidly than the material at the tip of the tail.
\end{itemize}
This results in a change of sign in the velocity gradient: radial velocities along the tail first decrease from the spiral to the tip and then increase at distances larger than $\sim$100--150~kpc. 
The projection chosen for our model is constrained by the observations: it makes this change of sign occur roughly at the location where the gas density peaks in the tidal debris.

Therefore the global kinematics of VirgoHI21 and its bridge can be simply explained as a result of streaming motions along a tidal structure combined with projection effects along the line-of-sight.

\section{Discussion}

\subsection{From tidal tails to fake dark galaxies}
\label{fakedark}
As shown in Section~\ref{sect:highvel}, tidal tails, especially those formed during high-velocity encounters, share many of the properties expected for dark galaxies: starless HI features, strong velocity gradients due in one case to streaming motions and in the other to the presence of a massive dark matter halo. If furthermore some gaseous condensations are present in the tail, they may resemble isolated dark galaxies: indeed the bridge to the parent galaxy can be very faint and hard or impossible to detect on moderately deep HI maps. 

Tidal tails do not necessarily have uniform profiles; denser regions can even lie in their outermost regions \citep[e.g.,][]{DBM04}. This can be because large pre-existing clouds from the parent disk are moved into the tail where they can form local overdensities \citep{elmegreen93}, or because some parts of an initially uniform tail condense under the effect of gravity \citep{BH92}. That these regions will lie far from the progenitor disk is a natural consequence of the extended flat rotation curves of spirals \citep{DBM04}. 
Depending on their location and mass, these denser parts of tails can even be self-gravitating, form stars and become finally independent objects with the mass of dwarf galaxies: the so-called Tidal Dwarf Galaxies \citep[TDGs,][for a review]{duc07}. In such cases, the internal motions within the young gravitationally bound object, in particular its rotation, will induce an additional velocity gradient.
On the other hand, the less massive condensations that have not reached the critical HI column density threshold to form stars will appear as detached HI clouds without any stellar counterpart.

Many of the known free-floating HI clouds, which are considered as Dark Galaxy candidates, have been found in clusters. In this environment, high velocity encounters have a high probability to occur, due to the large velocity dispersion of the cluster galaxies. Thus intra-cluster low-mass HI clouds of tidal origin could then be common but more dedicated studies should check this. Of course, this mechanism applies only if the parent galaxy had before the collision an extended HI disk. This requires in particular that it has not already crossed the cluster core where ram pressure would have contributed to truncate its gaseous disk. Tidal material generally quickly falls back onto the parent spiral, but this can take more than 2~Gyr in the outer parts \citep[e.g.,][]{BD06}. The cluster tidal field can even prevent tidal debris from falling back \citep{Mihos04}. This leaves time for fake dark galaxies of tidal origin to be observed while the interloper galaxy can be far away: at 1000~km~s$^{-1}$, it can be at a projected distance up to 2~Mpc two billion years later. 
\bigskip

\subsection{The nature of VirgoHI21}

After having argued that dark galaxies may in general be mistaken with tidal features and thus be fake ones, we discuss more specifically the nature of VirgoHI21, presenting pro and con arguments for the different scenarios proposed so far for its origin.

\subsubsection{Tidal debris?}

M07 argued that the VirgoHI21 + HI~bridge system cannot be a tidal tail from the spiral NGC~4254, because of the following reasons:
\begin{itemize}
\item Interacting galaxies generally have pair of tidal tails, and indeed the models in B05 have two-tailed morphologies, while NGC~4254 has no counter-tail.
\item The tidal HI clouds in B05 models are star-poor but not star-free, while HST optical observations show VirgoHI21 being completely dark.
\item The velocity gradient along the HI~bridge gets reversed around the VirgoHI21 cloud, which seems to rotate in a direction opposite to the rest of the HI bridge. This reversal is said by M07 not to be explained by interaction models as those of B05.
\item The velocity spread ($\Delta V=200$~km~s$^{-1}$) of VirgoHI21 could only be explained by an encounter at a comparable velocity (according to M07), implying that the interloper should not be further away than a few arcminutes and would have been identified. Moreover, low relative velocities are rare in the Virgo Cluster.
\item High velocity encounters are more common in this environment but velocities as large as $\sim$~1000~km~s$^{-1}$ are ``far too large to generate tidal features such as bridges and tails" (M07) .
\end{itemize}

If indeed VirgoHI21 is a genuine dark galaxy, as claimed by M07, the HI~bridge would then be a tail expulsed from the dark galaxy by an interaction with NGC~4254 or the cluster field.
We note here that our numerical model of VirgoHI21 which suggests that the tidal debris rather emanate from NGC~4254 addresses each of these above-mentioned concerns:
 
\begin{itemize}
\item The interaction occurred about 750~Myr ago, so that the counter-tail from the spiral is not necessarily expected to be observed anymore. By that time the interloper might also be far away.
\item The HI cloud formed in our model during a high--velocity encounter is free of old stars pre-existing to the galaxy interaction, and is not dense enough to form new stars.
\item The velocity spread ($\Delta V=200$~km~s$^{-1}$) of the HI structure does not imply that the galaxy interaction had a similar velocity. It is in fact accounted for by a high-velocity encounter.
\item The reversal of the velocity gradient along the tail is reproduced thanks to projection effects.
\end{itemize}

\begin{figure}
\centering
\includegraphics[width=\columnwidth]{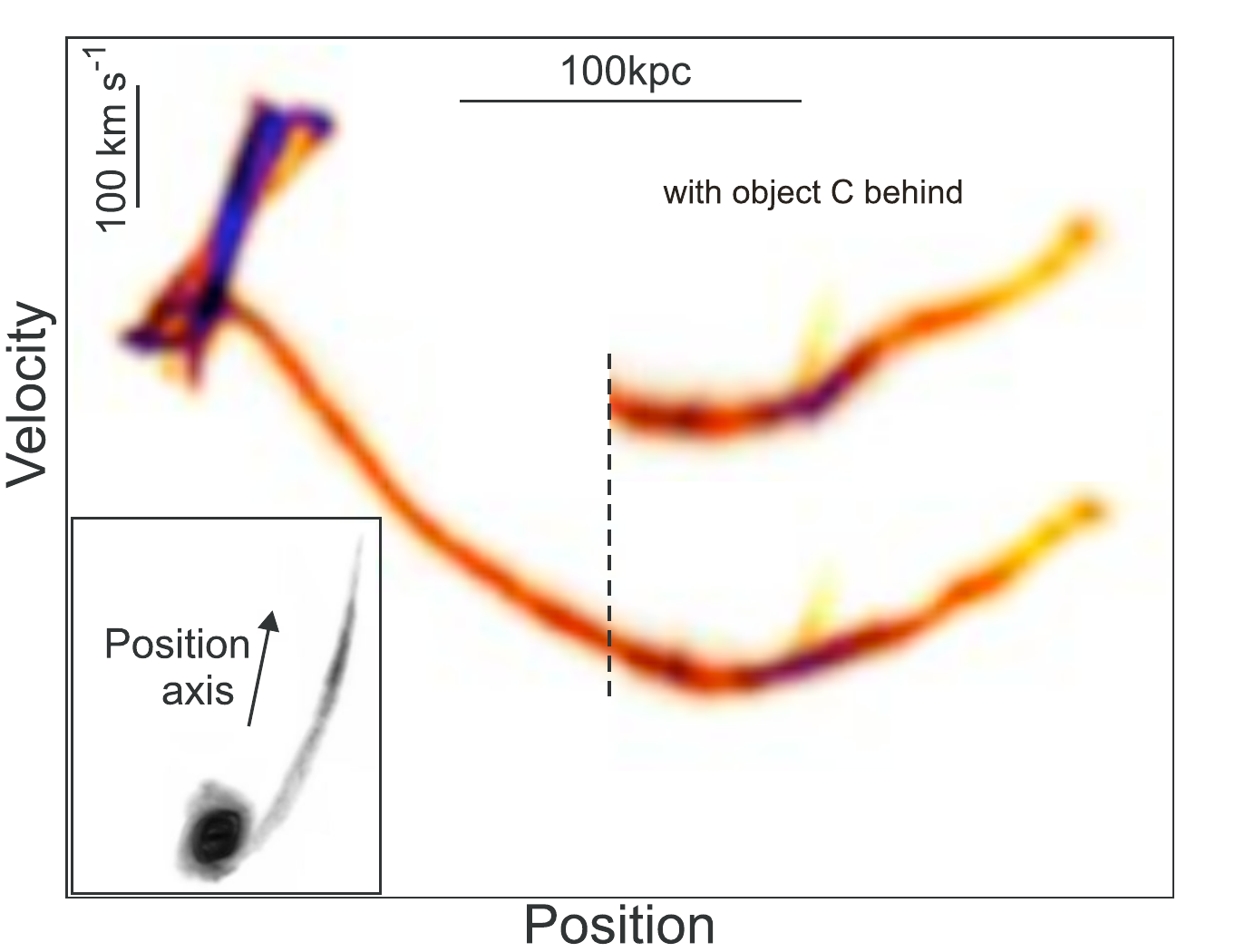}
\caption{Position--Velocity diagram of the gas for Model 7 at $t=$750 Myr when it best reproduces the actual morphology and velocity field of the system VirgoHI21/NGC 4254. The arrow indicates the axis of the ``Position" direction. The PV diagram of a model where ``Object C" has been artificially added during the simulation is also shown. }\label{fig:posvel}
\end{figure}

Whereas the global kinematics of VirgoHI21 and its bridge can be simply explained by streaming motions along a tidal structure, in detail, the model and the observations show some differences. They may be noted when comparing the observed (Fig.~\ref{fig:pv-vhi21}) and simulated (Fig.~\ref{fig:posvel}) Position--Velocity diagrams along the tail; in particular, as put forward by M07, the velocity gradient towards VirgoHI21 is locally larger within the HI cloud, while in our model it is not more enhanced at this location than further away near the tip of the tail.

This local difference is actually not a real concern for our scenario. First, part of the local amplification of the gradient can result from the self-gravity of the VirgoHI21 cloud itself, that could be somewhat denser than in our model. In particular, as recently shown by \citet{B07}, tidal debris may contain a significant dark component that has not been included in the simulations. 

Alternatively, the velocity field can be disturbed by objects in the neighborhood, for instance by the nearby dwarf galaxy, SDSS J121804.26+144510.4, also known as object~C (see Fig.~\ref{fig:vhi21} and M07). The nature of this dwarf is actually unclear (see Appendix). We suppose here that it is a pre-existing object, physically interacting with the system. Object~C has a visible HI mass of $2\times 10^7$~M$_\sun$. We included its possible influence in a simple model. We describe it as a dwarf spheroid with a Plummer profile, total mass of $4\times 10^8$~M$_\sun$ (to include the dark matter mass) and scale-length of 5~kpc. It is located 8~kpc East from VirgoHI21 on the sky plane, and we assume it lies 20~kpc below VirgoHI21 in the radial direction. Simulating the whole trajectory of this dwarf in the simulation from $t=0$ would introduce too many parameters. Since we just want to illustrate its possible local effect, we simply add it as a fixed mass in the late stage, linearly increasing its mass from $0$ at $t=600$~Myr to the final value at $t=700$. The velocity perturbation induced by object~C is shown on Figure~\ref{fig:posvel}. The model qualitatively reproduces the local amplification of the velocity gradient at the position of VirgoHI21, and in particular the ``S-shape" of the velocity profile visible in the Position-Velocity diagram (see Fig.~\ref{fig:pv-vhi21}). Tuning the parameters of the simulation may help to further reproduce the exact kinematical feature. Obviously other objects, such as NGC~4262 -- the gas--rich galaxy to the East in Figure~\ref{fig:vhi21} -- might have also crossed the trajectory of the tidal tail and slightly interacted with it. In any case, whatever the real explication is, the ``S-shape" of the velocity profile of VirgoHI21 is not inconsistent with a tidal origin.

\bigskip

A second critical issue is the identification of the interloper responsible for the collision.
In our high-velocity scenario, the interloper now lies at a projected distance of 400~kpc to the WNW of NGC~4254. A massive spiral is in fact present near this position: NGC~4192 (M~98). This galaxy, which is seen close to edge-on, has a maximum rotation velocity corrected for inclination  of 236~km~s$^{-1}$, significantly higher than that of NGC~4254 (193~km~s$^{-1}$, according to the HyperLeda database), making it compatible with the 1.5:1 mass ratio used in our simulation. In our model, the interloper is today approaching us with a radial velocity of 715~km~s$^{-1}$ w.r.t. the target spiral. In the real Universe, NGC~4192 is indeed approaching, but with a relative velocity of $\sim 2000$~km~s$^{-1}$ with respect to NGC~4254, i.e. much larger than in our model. This difference could be explained by the tidal field of the cluster which is not taken into account in our model. Adding it would introduce too many additional free parameters since the radial position and tangential velocity of the studied objects with respect to the cluster are unknown. The cluster gravitational field can modify some details in the properties of tidal tails in the long term \citep{Mihos04}, but not their fundamental characteristics. However the cluster tidal field can have a more dramatic effect on the relative orbit of the distant galaxy pair over the nearly 750~Myr period since the encounter. Indeed, a cluster with the mass typical of Virgo, $M_\mathrm{C}=10^{15}$~M$_\sun$, at a distance $R_\mathrm{C}=1$~Mpc from the pair, and a typical separation between the two galaxies $d\sim 300$~kpc (which is the average 3-D separation from the time of the encounter until today) would create a tidal acceleration
\begin{equation}
a_\mathrm{t} \simeq \frac{G M_\mathrm{C} d} {{R_\mathrm{C}} ^3}
\end{equation}
and, over a timescale of $T=750$~Myr, a relative velocity impulsion of
\begin{equation}
\Delta V_\mathrm{t} \simeq \frac{G M_\mathrm{C} d} {{R_\mathrm{C}} ^3} \times T
\end{equation}
which would come in addition to the relative velocity found in our model. The values above imply $\Delta V \simeq 1100$~km~s$^{-1}$. This additionnal velocity difference would account for the observed relative velocities of NGC~4192/4254, and goes in the right direction if NGC~4254 lies behind the Virgo Cluster center. 
 
NGC~4192 is thus a fully possible interloper, in spite of its large distance and relative velocity. This galaxy does not show a strongly disturbed HI disk on moderately deep HI maps \citep{Chung05}. One reason could be unfavorable internal/orbital parameters, for instance a retrograde orbit. Alternatively, faint tidal debris may be present, but only visible on deep HI observations, similar to those that revealed the existence of VirgoHI21.

However, we do not intent to claim that NGC 4192 is the only possible interloper. Other combination of orbits/projection/age can certainly reproduce the properties of VirogHI21, and a galaxy flying away at $\sim 1000$~km~s$^{-1}$ during $\sim 1$~Gyr can be at a projected distance of 1~Mpc today, possibly even at the center of the Virgo Cluster. The number of massive interloper candidates is then large, making hard to identify the real culprit. This is anyway not required for our demonstration that VirgoHI21 can be a tidal debris, since we have shown that possible interlopers do exist. 

\subsubsection{A kinematically decoupled Tidal Dwarf Galaxy?}
The presence of a strong velocity gradient in a tidal tail, if not due to streaming motions, may actually pinpoint the presence of a gravitationally bound object that need not be a pre-existing dark-matter dominated galaxy. 
Massive substructures in tidal tails often become kinematically decoupled, self-gravitating and form new stars, becoming rotating Tidal Dwarf Galaxies.  This is the case for VCC~2062, a  TDG candidate in Virgo \citep{Duc07b}. VirgoHI21 appears as a gas condensation within a tidal tail; it is currently not a TDG since it is starless, but could be its gaseous progenitor. 
Whether stars will be formed in this structure later is questionable. If the surface column density has remained unusually very low during the first several hundreds of Myr after the formation of VirgoHI21, it is unlikely that star-formation is ignited later on. 
On the other hand, the dynamical collapse time of the cloud, $1/\sqrt{G \rho}$  \citep{elmegreen02},  is as large as 400 -- 500 Myr for an initially resting system and an average volume mass density in our modeled cloud of $\rho \sim 10^{-3}$~M$_{\sun}$~pc$^{-3}$ (this is also about the density of the real VirgoHI21 cloud assuming a vertical scale-height of $\sim$ 300~pc). Comparing this time scale  to  the age of the cloud,   about 500~Myr (it appears in the model at $t=200-300$~Myr), one may conclude that  VirgoHI21 would barely have had the time to collapse and form stars even in the most favorable conditions and incidentally  that the absence of stars today is not dependent on an arbitrary choice of the threshold.  In other words, the system could still be contracting under the effect of its internal gravity today, and begin to form stars later-on {\it if} its density comes to exceed the star formation threshold.

However, the main argument against VirgoHI21 being {\it yet} a TDG fully responsible for the observed velocity gradient is the large dynamical mass inferred from the rotation curve: in galaxies made out of collisional debris, the dynamical mass should be of the same order as the luminous one, even if the presence of dark baryons may cause some differences between them \citep{B07}. In the case of VirgoHI21, the dynamical mass inferred from the velocity curve is more than a factor of 3000 greater than the luminous one, i.e. that of the HI component. Clearly, streaming motions provide a much more reasonable explanation for the kinematical feature observed near VirgoHI21, if indeed this object is of tidal origin.

\subsubsection{Harrasment by the cluster field?}
 In the group environment, \cite{bekki05b} proposed a scenario in which  the group tidal field is able to strip gas from HI-rich galaxies,  explaining the presence of isolated intergalactic HI clouds.  Following this idea, B05 presented a model in which the combined action of galaxy-galaxy interactions and the cluster tidal field produce debris with properties similar to Dark Galaxies. 
\citet{haynes07} even suggested that VirgoHI21 and the whole HI structure would result  from just the  long-term harassment by the large-scale cluster potential. However, the tidal field exerted by a structure of mass $M$ and typical scale $R$ scales as $M/R^3$. The tidal field of the Virgo Cluster ($10^{15}$~M$_\sun$, 1~Mpc) at the present distance of NGC~4254 is then more than ten times smaller than that of the interloper galaxy in our interaction model ($2\times 10^{12}$~M$_\sun$, 50~kpc). It is just unlikely that the cluster field can develop a tail as long as the galaxy interaction can do. The harassment process has a longer timescale than the galaxy pair interaction, but over long timescales the orientation changes, which hardly accounts for the single, thin and long tail around NGC~4254. This structure is more typical of a short and violent interaction like a close galaxy encounter than a weaker and longer process like the harassment by the global cluster field.

\subsubsection{Ram pressure stripping?}
\label{rampressure}

Ram pressure exerted by the cluster hot gas may also expulse gas from the outer regions of spiral disks and create isolated HI clouds without any optical counterpart. Yet, this scenario suffers fundamental concerns, also pointed out by M07, in particular:

\begin{itemize}
\item structures known to result from ram-pressure stripping are rather short and thick as observed in Virgo \citep{crowl05,lucero05,vollmer06,chung07} and suggested by hydrodynamical simulations \citep[e.g.,][]{Roediger07}, while the HI bridge of VirgoHI21 is much thinner and longer.
\item the kinematics is not directly explained too, in particular the reversing velocity gradient in the HI~bridge, which in the context of a tidal interaction results from streaming motions along a curved tail (in 3-D) more or less seen edge-on. This may also be the case for ram pressure but should be demonstrated by a model. 
\end{itemize}

\cite{vollmer05} put forward the role of ram pressure, combined with a tidal interaction,  in shaping the internal gas  distribution, with its $m=1$ structure,  and velocity field of NGC~4254. This partly explains  why,  in the innermost regions, the detailed kinematics of the spiral  presents   some differences with that of our model which did not take into account the intracluster medium. 
More recently, \cite{Kantharia07} presented a low radio frequency continuum map  of the galaxy which is best explained invoking a ram pressure scenario. However so far, its possible contribution on the properties of the HI bridge and VirgoHI21 has not yet been investigated. The two above-mentioned papers do not claim  that their origin is  ram-pressure, and indeed our pure tidal model is  able to reproduce these features provided that the HI disk was originally much more extended than the optical radius -- an hypothesis  which was not adopted in \cite{vollmer05}.

\subsubsection{A dark galaxy?}

Showing that virgoHI21 {\it can} be a tidal debris taking the appearance of a dark galaxy does not directly rule out the possibility that it is a real dark galaxy. The dark galaxy hypothesis however suffers several difficulties unexplained so far:

\begin{itemize}
\item the maximal velocity gradient is not centered on the peak of the HI emission, but actually lies on one side of the VirgoHI21 cloud. This is unexpected for a rotating HI disk within a massive dark halo. One may propose that the on-going interaction with NGC~4254 causes this asymmetry in the velocity field, but that NGC~4254 is massive and close enough to induce such major disturbances remains to be shown.
\item within the assumption that VirgoHI21 is a real dark galaxy, the HI bridge is a tidal tail expulsed from it and captured by the more massive galaxy NGC~4254 (M07). The gas in the HI bridge, falling onto NGC~4254 from 150~kpc away, is not expected to have the same velocity as the local gas settled in rotation: it could be on retrograde, polar or direct orbits but with different velocities. However, the base of the HI bridge has a radial velocity coherent with the outer disk of NGC~4254 to which it is morphologically connected: the velocity step between the base of the tidal tail and the outer disk is in fact less than $50$~km~$^{-1}$, much smaller than the circular velocity there \citep[see data in M07 and also][]{phookun93}. This further suggests that the HI material in the bridge comes from NGC~4254.
\end{itemize}

These facts are naturally explained by the tidal scenario proposed for VirgoHI21. Whether they can also be addressed with the Dark Galaxy hypothesis remains to be demonstrated, in particular with a numerical model.

Although challenged, the putative existence of Dark Galaxies as massive as VirgoHI21, has fostered a number of follow-up works. As discussed by \citet{karachentsev06} such invisible ghost objects should tidally perturb galaxies in their neighborhood, explaining why a fraction of apparently isolated spiral stellar disks seem to show signs of an external perturbation. We note however that other mechanisms may account for them, such as accretion of diffuse gas \citep{bournaud05m1}.

\section{Summary}

Using a series of numerical simulations, we have investigated the role of the initial impact velocity in the formation of tidal tails during galaxy-galaxy collisions. This work was motivated by the fact that collisional debris, which may become detached from their parent galaxies, exhibit many of the properties expected for the disputed class of ``dark galaxies", as initially suspected, among others, by Bekki et al. (2005). 
We found that, contrary to common belief, high-velocity fly-bys at 1000~km~s$^{-1}$ or more can generate the development of streams of tidal origin. These tails are less massive than those formed during lower-velocity encounters, but can be as long. An important difference is that tidal tails from high-velocity interactions contain a higher fraction of gas and can even be devoid of stars. Indeed the internal stellar disk of the parent galaxy is only weakly disturbed and the tails are mostly formed from the external HI disks. This explains why the role of high-speed collisions has so far been neglected, except to account for the harassment process in clusters of galaxies with repeated distant interactions.

Streaming motions are present in collisional debris and may generate in detached clouds velocity gradients comparable to those expected for rotating, self-gravitating, bodies. Starless gas clouds, showing apparent but fake signs of rotation, are thus the natural by-product of high-velocity collisions. Such characteristics are actually exactly those used to define an object as a Dark Galaxy candidate: an HI detection, without any stellar counterpart, and with a kinematics tracing the presence of a massive dark matter halo.

A candidate Dark Galaxy that has recently attracted much attention is VirgoHI21: an HI cloud in the Virgo Cluster, apparently connected by a faint HI bridge to the spiral galaxy NGC~4254, and exhibiting a strong velocity gradient of 200 km~s$^{-1}$ attributed to a dark matter halo as massive as 10$^{11}$ M$_{\sun}$ (Minchin et al. 2007). 
We propose here that this intriguing HI structure results simply from a high-velocity collision, a common phenomenon in clusters of galaxies. Our numerical simulation reproduces the morphology and kinematics of the whole system, including the parent spiral galaxy, believed to be NGC~4254, the HI bridge, and the VirgoHI21 cloud. A counter-tidal tail was formed too, but was shorter and has now fallen back onto the parent spiral galaxy, explaining the present single-tailed morphology of NGC~4254. This model assumes that the interaction occurred 750~Myr ago with a massive galaxy, nowadays lying 400~kpc away in projected distance. A candidate for the interloper is the spiral NGC~4192 but other galaxies/orbits are probably possible too, seen the large number of massive galaxies within 1~Mpc is this region. The concerns raised by Minchin et al. (2007) against the tidal scenario were all addressed. 

With the availability of deep HI surveys, the number of starless free floating HI clouds and thus of Dark Galaxy candidates might increase. Objects as spectacular as VirgoHI21 for which a tidal origin might easily be assessed by observations and numerical simulations are extremely rare. HI maps may not be sensitive enough to detect the HI bridge linking low-mass tidal debris to their parent galaxies. They may also be old and have already lost their umbilical cord, a process accelerated in the cluster environment. In such conditions, proving unambiguously a tidal origin may become more difficult. If the cloud is self-gravitating, its dynamical mass may be derived provided that the spatial and spectral resolutions of the HI maps are high enough. If this total mass is comparable to the luminous one, actually the HI mass, the absence of a dark component excludes the Dark Galaxy hypothesis. A measure of the metallicity of the HI cloud, possible if it contains HII regions or has by chance a background quasar in its line of sight, would also provide a reliable test: tidal debris are made of metal-rich, pre-enriched, material while genuine Dark Galaxies would be made of pristine, metal-poor, gas. The detection of molecular gas, in particular the metallicity sensitive CO millimetric line, would also reveal a pre-enrichment consistent with a tidal origin.

In any cases, numerical simulations of high velocity collisions indicate that tidal debris corresponding to fake dark galaxies could be numerous even in dense environments like the Virgo Cluster. Whether genuine dark galaxies exist then remains to be proven.

\acknowledgments

The simulations in this paper were performed on the NEC-SX8R and SX8 vectorial computers at the CEA/CCRT and CNRS/IDRIS computing centers. We acknowledge the usage of the HyperLeda database. We are particularly grateful to Robert Minchin and Brian Kent who sent us the fully reduced WSRT and Arecibo HI datacubes respectively, that we used in this paper to constrain the numerical models. This work has largely benefited from stimulating discussions with Elias Brinks, Bernd Vollmer, Riccardo Giovanelli, Martha Haynes, Josh Simon, Jonathan Davies, Robert Minchin and Mike Disney.

\appendix
\section{The strange nature of Object C}
The so-called Object C -- SDSS J121804.26+144510.4 -- is a gas--rich dwarf galaxy lying very near VirgoHI21, to the East (see Fig.~1). Its velocity is offset by just 100~km~s$^{-1}$ (see Fig.~2). This proximity suggests that both objects might belong to the same HI structure. However, whereas the Dark Galaxy candidate has no optical counterpart, object C is associated to an optically bright component, with an SDSS r--band magnitude of 16.6. The presence of stars in it may help to determine its age, chemical properties and nature and thus by extrapolation to constrain the origin of the whole HI structure. 

An optical spectrum of the dwarf galaxy is available in the Sloan database. It exhibits emission lines indicative of an on-going star-formation activity consistent with its blue color. Oxygen abundances derived in its HII regions with three different empirical methods give contradictory results, with 12+log(O/H) either equal to 8.0, 8.3 or even 8.5 (V\'{\i}lchez, J.~M. and Iglesias-P\`{a}ramo, J., private communication, and our own measurement). The low values would be consistent with the idea that the object is a classical pre-existing star-forming dwarf while the high values would suggest it is made of material which had been pre-enriched in another galaxy, as for Tidal Dwarf Galaxies. Would Object C be a TDG, then most likely the nearby HI bridge and VirgoHI21 would also be tidal debris that contrary to the former would not have managed to form stars. Clearly follow-up observations would be required to disclose its real nature.   In particular obtaining near--infrared data and combining them with the already available GALEX/UV and optical data  would constrain the age of the stellar population; a millimetric CO spectrum that traces its molecular gas content would indirectly probes its metallicity. 

In any cases, the discrepancies between the various estimates of the oxygen abundance reveal very unusual line ratios in the optical spectrum of Object C with respect to other star-forming dwarfs in Virgo \citep{Vilchez03}; these peculiarities still need to be understood.


\begin{thebibliography}{43}
\expandafter\ifx\csname natexlab\endcsname\relax\def\natexlab#1{#1}\fi

\bibitem[{{Barnes} \& {Hernquist}(1992)}]{BH92}
{Barnes}, J.~E. \& {Hernquist}, L. 1992, \nat, 360, 715

\bibitem[{{Bekki} {et~al.}(2005a){Bekki}, {Koribalski}, \& {Kilborn}}]{bekki05}
{Bekki}, K., {Koribalski}, B.~S., \& {Kilborn}, V.~A. 2005a, \mnras, 363, L21 (B05)

\bibitem[{{Bekki} {et al.}(2005b)}]{bekki05b} Bekki, K., Koribalski, 
B.~S., Ryder, S.~D., \& Couch, W.~J.\ 2005b, \mnras, 357, L21 


\bibitem[{{Bournaud} \& {Combes}(2002)}]{BC02}
{Bournaud}, F. \& {Combes}, F. 2002, \aap, 392, 83

\bibitem[{{Bournaud} \& {Combes}(2003)}]{BC03}
---. 2003, \aap, 401, 817

\bibitem[{{Bournaud} {et~al.}(2003)}]{BDM03}
Bournaud, F., Duc, P.-A., \& Masset, F. 2003, A\&A, 411, L469

\bibitem[{{Bournaud} {et~al.}(2005){Bournaud}, {Combes}, {Jog}, \&
 {Puerari}}]{bournaud05m1}
{Bournaud}, F., {Combes}, F., {Jog}, C.~J., \& {Puerari}, I. 2005, \aap, 438,
 507

\bibitem[{{Bournaud} \& {Duc}(2006)}]{BD06}
{Bournaud}, F. \& {Duc}, P.-A. 2006, \aap, 456, 481

\bibitem[{{Bournaud} {et~al.}(2004){Bournaud}, {Duc}, {Amram}, {Combes}, \&
 {Gach}}]{bournaud04}
{Bournaud}, F., {Duc}, P.-A., {Amram}, P., {Combes}, F., \& {Gach}, J.-L. 2004,
 \aap, 425, 813

\bibitem[{{Bournaud} {et~al.}(2007){Bournaud}, {Duc}, {Brinks}, {Boquien},
 {Amram}, {Lisenfeld}, {Koribalski}, {Walter}, \& {Charmandaris}}]{B07}
{Bournaud}, F., et al. 2007, Science, 316, 1166


\bibitem[{{Carignan} \& {Freeman}(1988)}]{carignanfreeman88}
{Carignan}, C. \& {Freeman}, K.~C. 1988, \apjl, 332, L33

\bibitem[{{Chung} {et~al.}(2005)}]{Chung05}
{Chung}, A., {van Gorkom}, J.~H., {Kenney}, J.~D.~P., \& {Vollmer}, B. 2005, in
 ASP Conference Series, Vol. 331, Extra-Planar
 Gas, 275 (arXiv:astro--ph/0507592)

\bibitem[{{Chung} {et~al.}(2007)}]{chung07}
{Chung}, A., {van Gorkom}, J.~H., {Kenney}, J.~D.~P., \& {Vollmer}, B. 2007,
 \apjl, 659, L115

\bibitem[{{Crowl} {et~al.}(2005){Crowl}, {Kenney}, {van Gorkom}, \&
 {Vollmer}}]{crowl05}
{Crowl}, H.~H., {Kenney}, J.~D.~P., {van Gorkom}, J.~H., \& {Vollmer}, B. 2005,
 \aj, 130, 65

\bibitem[{{Davies} {et~al.}(2004){Davies}, {Minchin}, {Sabatini}, {van Driel},
 {Baes}, {Boyce}, {de Blok}, {Disney}, {Evans}, {Kilborn}, {Lang}, {Linder},
 {Roberts}, \& {Smith}}]{davies04}
{Davies}, J., et al. 2004, \mnras, 349, 922

\bibitem[{{Davies} {et~al.}(2006){Davies}, {Disney}, {Minchin}, {Auld}, \&
 {Smith}}]{davies06}
{Davies}, J.~I., {Disney}, M.~J., {Minchin}, R.~F., {Auld}, R., \& {Smith}, R.
 2006, \mnras, 368, 1479

\bibitem[{{de Blok} {et~al.}(2005){de Blok}, {Walter}, {Brinks}, {Thornley}, \&
 {Kennicutt}}]{things}
{de Blok}, W.~J.~G., {Walter}, F., {Brinks}, E., {Thornley}, M.~D., \&
 {Kennicutt}, Jr., R.~C. 2005, in Astronomical Society of the Pacific
 Conference Series, Vol. 329, Nearby Large-Scale Structures and the Zone of
 Avoidance, ed. A.~P. {Fairall} \& P.~A. {Woudt}, 265

\bibitem[{{Doyle} {et~al.}(2005)}]{doyle05}
{Doyle}, M.~T., et al. 2005, \mnras, 361, 34


\bibitem[{{Dubinski} {et~al.}(1996)}]{dubinski1996}
{Dubinski}, J., {Mihos}, J.~C., \& {Hernquist}, L., 1996, ApJ, 462, 576


\bibitem[{{Duc} {et~al.}(2007a){Duc}, {Bournaud}, \& {Boquien}}]{duc07}
{Duc}, P.-A., {Bournaud}, F., \& {Boquien}, M. 2007, in IAU Symposium, Vol.
 237, IAU Symposium, ed. B.~G. {Elmegreen} \& J.~{Palous}, 323
 (arXiv:astro--ph/0610047)

\bibitem[Duc et al.(2007b)]{Duc07b} Duc, P.-A., Braine, J., 
Lisenfeld, U., Brinks, E., \& Boquien, M.\ 2007, A\&A in press (arXiv:0709.2733) 


\bibitem[{{Duc} {et~al.}(2004){Duc}, {Bournaud}, \& {Masset}}]{DBM04}
{Duc}, P.-A., {Bournaud}, F., \& {Masset}, F. 2004, \aap, 427, 803

\bibitem[{{Eneev} {et~al.}(1973)}]{eneev73}
{Eneev}, T.~M., {Kozlov}, N.~N., \& {Sunyaev}, R.~A., 1973, A\&A, 22, 41
  

\bibitem[{{Elmegreen} {et~al.}(1993){Elmegreen}, {Kaufman}, \&
 {Thomasson}}]{elmegreen93}
{Elmegreen}, B.~G., {Kaufman}, M., \& {Thomasson}, M. 1993, \apj, 412, 90

\bibitem[Elmegreen(2002)]{elmegreen02} Elmegreen, B.~G.\ 2002, 
\apj, 577, 206 


\bibitem[{{Giovanelli} {et~al.}(2007){Giovanelli}, {Haynes}, {Kent},
 {Saintonge}, {Stierwalt}, {Altaf}, {Balonek}, {Brosch}, {Brown}, {Catinella},
 {Furniss}, {Goldstein}, {Hoffman}, {Koopmann}, {Kornreich}, {Mahmood},
 {Martin}, {Masters}, {Mitschang}, {Momjian}, {Nair}, {Rosenberg}, \&
 {Walsh}}]{alfalfa1}
{Giovanelli}, R., et al. 2007, \aj, 133, 2569


\bibitem[{{Guiderdoni}(1987)}]{guiderdoni87}
{Guiderdoni}, B., 1987, A\&A, 172, 27


\bibitem[{{Haynes} {et~al.}(2007){Haynes}, {Giovanelli}, \& {Kent}}]{haynes07}
{Haynes}, M.~P., {Giovanelli}, R., \& {Kent}, B.~R. 2007, \apj, 665, L19

\bibitem[{{Hibbard} {et~al.}(1994){Hibbard}, {Guhathakurta}, {van Gorkom}, \&
 {Schweizer}}]{hibbard94}
{Hibbard}, J.~E., {Guhathakurta}, P., {van Gorkom}, J.~H., \& {Schweizer}, F.
 1994, \aj, 107, 67

\bibitem[Kantharia et al.(2007)]{Kantharia07} Kantharia, N.~G., 
Pramesh Rao, A., \& Sirothia, S.~K.\ 2007, MNRAS in press (arXiv:0709.4532) 


\bibitem[{{Karachentsev} {et~al.}(2006){Karachentsev}, {Karachentseva}, \&
 {Huchtmeier}}]{karachentsev06}
{Karachentsev}, I.~D., {Karachentseva}, V.~E., \& {Huchtmeier}, W.~K. 2006,
 \aap, 451, 817

\bibitem[{{Kennicutt}(1998)}]{kennicutt98}
{Kennicutt}, Jr., R.~C. 1998, \apj, 498, 541

\bibitem[{{Kent} {et~al.}(2007){Kent}, {Giovanelli}, {Haynes}, {Saintonge},
 {Stierwalt}, {Balonek}, {Brosch}, {Catinella}, {Koopmann}, {Momjian}, \&
 {Spekkens}}]{alfalfa2}
{Kent}, B.~R., et al. 2007, \apj, 665, L15

\bibitem[{{Lucero} {et~al.}(2005){Lucero}, {Young}, \& {van Gorkom}}]{lucero05}
{Lucero}, D.~M., {Young}, L.~M., \& {van Gorkom}, J.~H. 2005, \aj, 129, 647

\bibitem[{{Martin} \& {Kennicutt}(2001)}]{martinkenn01}
{Martin}, C.~L. \& {Kennicutt}, Jr., R.~C. 2001, \apj, 555, 301

\bibitem[{{Meyer} {et~al.}(2004)}]{meyer04}
{Meyer}, M.~J., et al. 2004, \mnras, 350, 1195

\bibitem[{{Mihos}(2004)}]{Mihos04}
{Mihos}, C. 2004, in IAU Symposium 217, Recycling intergalactic and
 interstellar matter, ed. P.~A. Duc, J.~Braine, \& E.~Brinks (ASP), 390
 (arXiv:astro--ph/0401557)

\bibitem[{{Minchin} {et~al.}(2005){Minchin}, {Davies}, {Disney}, {Boyce},
 {Garcia}, {Jordan}, {Kilborn}, {Lang}, {Roberts}, {Sabatini}, \& {van
 Driel}}]{minchin05}
{Minchin}, R., et al. 2005, \apjl, 622, L21

\bibitem[{{Minchin} {et~al.}(2007){Minchin}, {Davies}, {Disney}, {Grossi},
 {Sabatini}, {Boyce}, {Garcia}, {Impey}, {Jordan}, {Lang}, {Marble},
 {Roberts}, \& {van Driel}}]{minchin07}
{Minchin}, R., et al. 2007, ApJ, in press (arXiv:706.1586) (M07)

\bibitem[{{Moore} {et~al.}(1996){Moore}, {Katz}, {Lake}, {Dressler}, \&
 {Oemler}}]{moore96}
{Moore}, B., {Katz}, N., {Lake}, G., {Dressler}, A., \& {Oemler}, A. 1996,
 \nat, 379, 613

\bibitem[{{Phookun} {et~al.}(1993){Phookun}, {Vogel}, \& {Mundy}}]{phookun93}
{Phookun}, B., {Vogel}, S.~N., \& {Mundy}, L.~G. 1993, \apj, 418, 113

\bibitem[{{Roberts} \& {Haynes}(1994)}]{robertshaynes94}
{Roberts}, M.~S., \& {Haynes}, M.~P., 1994, \araa, 32, 115

\bibitem[{{Roediger} \& {Brueggen}(2007)}]{Roediger07}
{Roediger}, E. \& {Brueggen}, M. 2007, \mnras, 380, 1399 

\bibitem[{{Simon} {et~al.}(2006){Simon}, {Blitz}, {Cole}, {Weinberg}, \&
 {Cohen}}]{simon06}
{Simon}, J.~D., {Blitz}, L., {Cole}, A.~A., {Weinberg}, M.~D., \& {Cohen}, M.
 2006, \apj, 640, 270

\bibitem[{{Simon} {et~al.}(2004){Simon}, {Robishaw}, \& {Blitz}}]{simon04}
{Simon}, J.~D., {Robishaw}, T., \& {Blitz}, L. 2004, in Astronomical Society of
 the Pacific Conference Series, Vol. 327, Satellites and Tidal Streams, ed.
 F.~{Prada}, D.~{Martinez Delgado}, \& T.~J. {Mahoney}, 32

\bibitem[{{Taylor} \& {Webster}(2005)}]{taylor05}
{Taylor}, E.~N. \& {Webster}, R.~L. 2005, \apj, 634, 1067

\bibitem[{{Toomre}(1963)}]{toomre63}
{Toomre}, A. 1963, \apj, 138, 385

\bibitem[{{Tully}(2005)}]{tully05}
{Tully}, R.~B. 2005, \apj, 618, 214

\bibitem[{{van den Bosch} {et~al.}(2003){van den Bosch}, {Yang}, \&
 {Mo}}]{vandenbosch03}
{van den Bosch}, F.~C., {Yang}, X., \& {Mo}, H.~J. 2003, \mnras, 340, 771

\bibitem[{{V{\'{\i}}lchez} \& {Iglesias-P\'aramo}(2003)}]{Vilchez03}
{V{\'{\i}}lchez}, J.~M. \& {Iglesias-P\'aramo}, J. 2003, \apjs, 145, 225

\bibitem[{{Vollmer} {et~al.}(2005){Vollmer}, {Huchtmeier}, \& {van
 Driel}}]{vollmer05}
{Vollmer}, B., {Huchtmeier}, W., \& {van Driel}, W. 2005, \aap, 439, 921

\bibitem[{{Vollmer} {et~al.}(2006){Vollmer}, {Soida}, {Otmianowska-Mazur},
 {Kenney}, {van Gorkom}, \& {Beck}}]{vollmer06}
{Vollmer}, B., {Soida}, M., {Otmianowska-Mazur}, K., {Kenney}, J.~D.~P., {van
 Gorkom}, J.~H., \& {Beck}, R. 2006, \aap, 453, 883

\end{thebibliography}

\end{document}